\documentclass[%
twocolumn,
 amsmath,amssymb,
 aps,
prstab,
]{revtex4}

\usepackage{graphicx}

\def\be{\begin{equation}}
\def\ee{\end{equation}}
\def\bea{\begin{eqnarray}}
\def\eea{\end{eqnarray}}

\def\St{{\rm St}}

\begin{document}

\title{Maximal air bubble entrainment at liquid drop impact}

\author{ Wilco Bouwhuis}
\author{Roeland C.A. van der Veen}
\author{Tuan Tran}
\author{Diederik L. Keij}
\author{Koen G. Winkels}
\author{Ivo R. Peters}
\author{Devaraj van der Meer}
\author{Chao Sun}
\author{Jacco H. Snoeijer}
\author{Detlef Lohse}

\affiliation{Physics of Fluids Group, Faculty of Science and Technology, Impact \& MESA+ Institutes, and Burgers Center for Fluid Dynamics,\\ University of Twente, 7500AE Enschede, The Netherlands}

\date{\today}


\begin{abstract}
At impact of a liquid drop on a solid surface an air bubble can be entrapped. 
Here we show that
two competing effects minimize the (relative) size of this entrained air bubble: For large drop impact velocity and large droplets the inertia of the liquid flattens the entrained bubble, whereas for small impact velocity and small droplets capillary forces minimize the entrained bubble.
However, we demonstrate experimentally, theoretically, and numerically that 
in between there is an optimum, leading to maximal air bubble entrapment. 
Our results have a strong bearing on various applications in 
printing technology, microelectronics, 
immersion lithography, diagnostics, or agriculture.
\end{abstract}

\maketitle

\begin{figure*}
 \includegraphics[width=14.0cm,angle=-0]{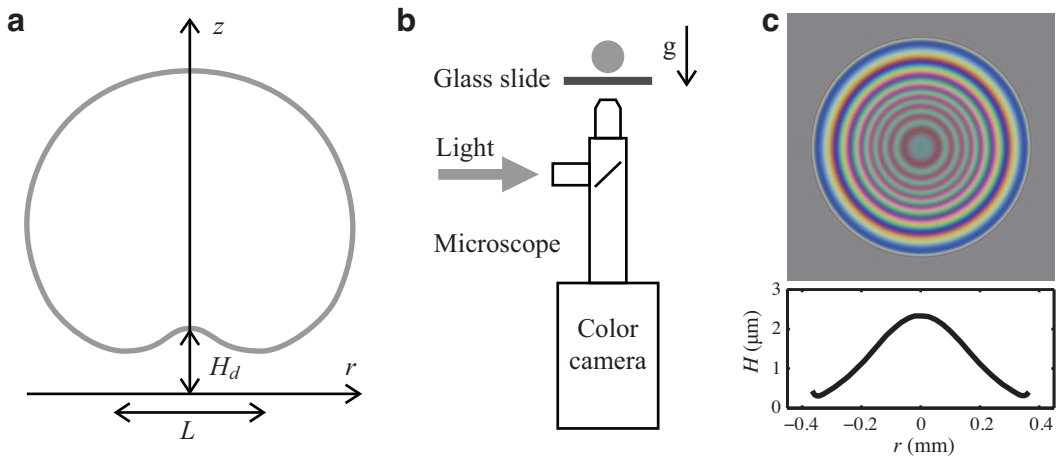}                    
    \caption{Experimental characterization of air bubble entrapment. 
(a) Sketch of dimple formation (not drawn to scale) just prior to impact. 
(b) Schematic of the experimental setup used to study droplet impact on smooth surfaces. 
An ethanol droplet  of typical radius $R = 0.9\,$mm 
falls on a glass slide of average roughness 10 nm. The impact velocity is varied by varying the falling height of
the droplet. For very small velocities below $0.31\,$m/s, the droplet is fixed at the tip of 0.4$\,$mm\--diameter capillary that is vertically translated downwards at a constant velocity. 
The bottom view is captured by a high-speed color camera (SA2, Photron Inc.). The camera is connected to a long working-distance microscope and a 5$\times$ objective to obtain a 2$\,$mm field of view. (c) An example of an interference pattern and the extracted air thickness profile. Note the difference in horizontal and vertical length scales.}
\label{fig1}
\end{figure*}

The impact of liquid droplets on surfaces is omnipresent in nature and technology, ranging from falling raindrops to 
applications in agriculture and inkjet printing.
The crucial question often is: 
How well does the  liquid wet a surface? The traditional view is that it is the surface tension which gives a quantitative answer. 
However, it has been shown recently that an air bubble can be entrapped under a liquid drop as it impacts on the surface 
\cite{dam04,tho05,man09,dri11,kol12,man12}. 
Also 
Xu \emph{et al.} \cite{xu05,xu07} revealed the important role of the surrounding air on the impact dynamics,
 including a possible splash formation. 
 The mechanism works as follows
 \cite{man09,dri11,kol12,man12}: The air between the falling drop and the surface is strongly squeezed, leading to a pressure buildup in the air under the drop. The enhanced
 pressure results in a dimple formation in the droplet and eventually to the entrapment of an air bubble (figure \ref{fig1}a). The very simple question we ask and answer in this paper is: For which impact velocity is the entrapped bubble maximal? 

%
Our experimental setup is shown in figure \ref{fig1}b and is similar to that of ref.\ \cite{vee12} where it is described in detail. An ethanol drop impacts on a smooth glass surface after detaching from a needle, or for velocities smaller than $0.32\,$m/s, after moving the needle downwards using a linear translation stage. A high-speed side view recording is used to measure the drop diameter and velocity.
A synchronized bottom view recording by a high-speed color camera is used to measure the deformed shape of the liquid drop. Colored interference patterns are created by high-intensity coaxial white light, which reflects from both the glass surface and the bottom of the droplet. Using a color-matching approach in combination with known reference surfaces, the complete air thickness profile can be extracted (shown in figure\,\ref{fig1}c).
For experiments done at larger impact velocities ($U>0.76\,$m/s),  we use a pulse of diffused laser light triggered by an optical switch. The thickness of the air film at the rim is assumed to be zero, and the complete air thickness profile can then be obtained from the monochromatic fringe pattern.
From these measurements we can determine the dimple height, $H_d$, and the volume of the entrained bubble, $V_b$, at the very moment of impact. This moment is defined by the first wetting of the surface, i.e., the moment when the concentric symmetry of the interference rings is lost. To calculate the bubble volume $V_b$, we integrate the thickness profile of the air layer trapped beneath the drop.
Alternatively, we can also measure the volume of the trapped bubble after impact when the liquid already wets the surface. Both measurements provide the same results. In the present article, we use the first approach.

\begin{figure*}
 \includegraphics[width=17.0cm,angle=-0]{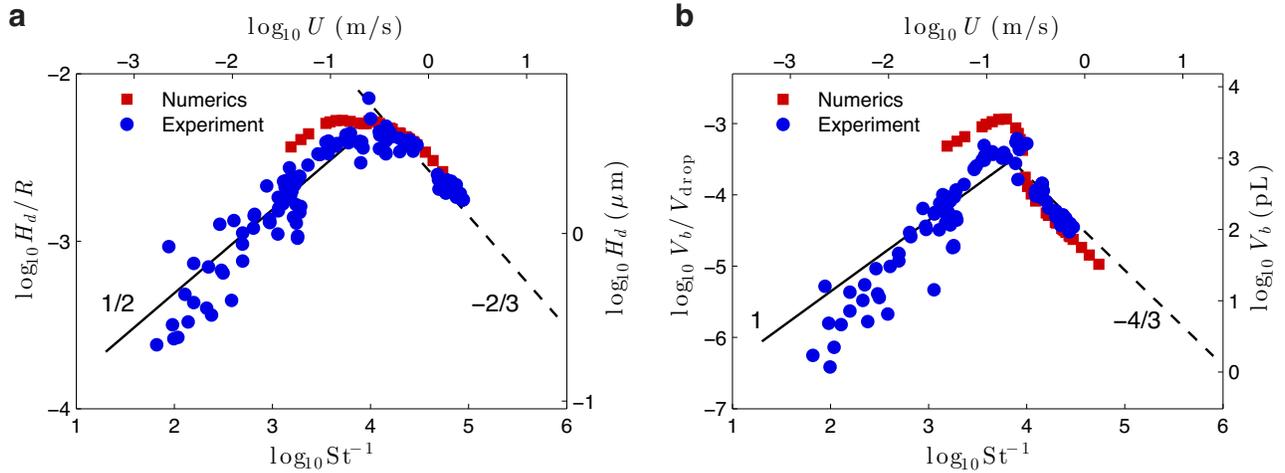}       
    \caption{Maximum entrapment of air bubbles. 
    (a) Dimple height $H_d$ and (b) entrained bubble volume $V_b$ as functions of the impact velocity $U$ (upper axes) and 
Stokes number $St$ (lower axes). The shape of the air layer can be characterized by the dimple height $H_d$ and the lateral extension $L$. 
Blue circles correspond to high-speed color interferometry measurements, red squares correspond to numerical simulations.
The straight lines correspond to the derived scaling laws in the capillary regime (solid) and inertial regime (dashed) with the respective
scaling exponents. The fitted prefactors for the scaling law of the dimple height are order unity, 0.3 and 3 respectively for the capillary and inertial regimes. The prefactors for the scaling laws of the volume, being the third power of length scales,
are larger, namely 9 and 170, respectively.
}
\label{fig2}
\end{figure*}

The results are shown in figure\ \ref{fig2}. Clearly, both dimple height at impact and the size of the entrained bubble 
have a pronounced maximum as function of the impact velocity $U$. The corresponding impact velocity for which the air entrainment is optimal is $U_o=0.25\,$m/s for an ethanol droplet of radius 
$R =0.9\,$mm (or the Stokes number $St_o = 1\times 10^{-4}$). 
While length scales are given in multiples
of the droplet radius $R$, following Brenner and coworkers \cite{man09,man12}
we express the impact velocity $U$ in terms of the (inverse) Stokes number $St$, defined with the dynamic air viscosity 
$\eta_g$ and the liquid density $\rho_l$
as $St= \eta_g/ (\rho_l R U) = \rho_g/ \rho_l Re^{-1}$, where $Re= \rho_g R U/\eta_g$ is the standard Reynolds number.
A further relevant parameter of the system is the surface tension $\gamma$, which can be  expressed in terms of the Weber 
number $We = \rho_l R U^2/\gamma$ or in terms of the capillary number $Ca = \eta_g U/\gamma = St \cdot We$. 


We compare and supplement our experimental findings on the dimple height at impact and the
entrained bubble size to numerical results. The numerics consists of an axisymmetric boundary integral (BI) simulation
for the liquid droplet (i.e., the droplet is assumed to obey potential flow), 
coupled to a lubrication approximation of the Stokes equation
\be
\frac{\partial P_g}{\partial r} \sim \, \eta_g \frac{\partial^2 u_r}{\partial y^2},
\label{stokes}
\ee
that describes the viscous, incompressible gas flow under the droplet \cite{man09,man10, hic10,hic11, egg10}. Here, $P_g(r,t)$ is the gas pressure, while $u_r$ is the radially outward velocity in the gas parallel to the surface (figure \ref{fig1}a). 
Note that the gas flow under the droplet is indeed viscous: An upper bound for the Reynolds number relevant for the lubrication flow gives $UH_d/\nu_g \sim 0.1$ for the highest impact velocity, and is typically much smaller for most of our experiments. 

We now give  more details on the numerical simulation:
The velocity field inside the droplet is described with a scalar velocity potential $\phi$, obeying the Laplace equation $\nabla^2 \phi = 0$. The axisymmetric droplet contour is described using cylindrical coordinates $r, z$ and is solved numerically by using the  BI method; the simulations are based on the numerical code described by \cite{ogu93,ber09,gekle11}. This BI simulation is an alternative way of solving the system of equations, compared to the method applied by \cite{man10}, in which case a Hilbert transform method was applied. In contrast to \cite{egg10}, we do not solve the complete Navier-Stokes equations, but do include dynamics of the air layer below the drop.
The dynamic boundary condition on the droplet contours is given by the unsteady Bernoulli equation, 
\begin{equation}
\left(\frac{\partial \phi}{\partial t}+\frac{1}{2}\left|\nabla \phi\right|^2\right)=-gz-\frac{\gamma}{\rho_l}\kappa(r,t)-\frac{P_{g}(r,t)-P_{\infty}}{\rho_l}.
\label{eq:unsBer}
\end{equation}
Here $t$ is time, $g$ the acceleration of gravity, $z$ the absolute height,  $\kappa(r,t)$ the interface curvature, 
and $P_{\infty}$ the far-field pressure. The key dynamical quantities in (\ref{eq:unsBer}) are the gas pressure $P_g(r,t)$ and the interface curvature $\kappa(r,t)$. The curvature is related to the dimple profile $H(r,t)$  by the geometric relation
\begin{equation}
\kappa(r,t) = \frac{\frac{\partial^2 H(r,t)}{\partial r^2}}{\left(1+\left(\frac{\partial H(r,t)}{\partial r}\right)^2\right)^{3/2}}+
\frac{\frac{\partial H(r,t)}{\partial r}}{r\left(1+\left(\frac{\partial H(r,t)}{\partial r}\right)^2\right)^{1/2}}.
\end{equation}
To close the problem, an additional equation is provided by the lubrication approximation for the viscous 
gas flow at the bottom of the droplet,
\begin{equation}
\frac{\partial H(r,t)}{\partial t} - \frac{1}{r}\frac{\partial}{\partial r}\left[\frac{r \left(H(r,t)\right)^3}{12\eta_g}\frac{\partial P_g(r,t)}{\partial r}\right] =0,
\label{eq:lubr}
\end{equation}
with boundary condition $P_g|_{r=R}=P_{\infty}$; the gas pressure at the top of the droplet is set to atmospheric.
 Contrarily to \cite{man10}, we do not incorporate effects of compressibility of the gas, since, following the analysis of \cite{hic11}, there is little influence of compressibility in the regime that is studied here. 
The initial conditions for the simulations consist of a spherical droplet with radius $R$ with a downward velocity $U$. The initial height is taken sufficiently high for the viscous lubrication pressure to be still negligible ($\sim10~\mathrm{\mu m}$). The number of nodes on the droplet surface for which the BI equations are solved is of order 100, with node density increasing for $r \rightarrow 0$. The number of nodes and the size of the time steps vary during the simulation, as a function of the local gap height and velocity of the droplet contour. The size of a time step is of order $10~\mathrm{ns}$. For any number of nodes, the coupling between gap height and pressure profile breaks down for some small value of $H$, since the pressure diverges at vanishing thickness of the air layer.
Consistent with the experimental resolution we continue our simulations until the minimum gap thickness reaches $0.4~\mathrm{\mu m}$, while ensuring that our algorithm remains accurate. This is the moment at which the values for $H_d$ and $V_b$ are extracted.

\begin{figure}
\centerline{\includegraphics[width=10cm]{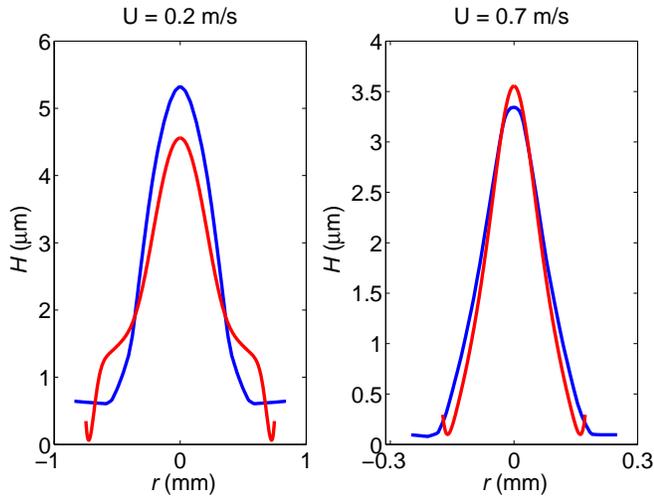}}
\caption{Comparison of experimental (blue) and numerical (red) dimple profiles for two different impact velocities; $U$=$0.2~\mathrm{m/s}$ ($\St$=$1.28\cdot10^{-4}$; crossover regime) and $U$=$0.7~\mathrm{m/s}$ ($\St$=$3.66\cdot10^{-5}$; inertial regime).
}
\label{fig:profcomp}
\end{figure}

The results of the numerical calculations are shown in figure \ref{fig2}, together with the experimental data,
showing very good agreement with our experimental results: in particular, we observe the pronounced maxima in the dimple size and in the entrained bubble volume at the optimal Stokes number $St_o$. In the numerically obtained bubble volume, we observe a jump exactly at the crossover regime. This jump originates from a change in the shape of the dimple. Figure \ref{fig:profcomp} compares the experimental and numerical dimple profiles for an impact velocity at the crossover regime ($U$=$0.2~\mathrm{m/s}$) and an impact velocity in the inertial regime ($U$=$0.7~\mathrm{m/s}$). While the profiles are in excellent agreement in the inertial regime (both volume and dimple height), the numerical profile develops a ``double dimple'' at the lower impact speed. This variation in shape results in the jump observed for the numerical bubble volumes in the crossover regime (see figure \ref{fig2}b). In all cases, however, the dimple height $H_d$ is in quantitative agreement without any adjustable parameters.

%

Numerical and experimental results together suggest 
scaling laws $H_d/R \sim St^{2/3} $ and $V_b/R^3 \sim St^{4/3}$ for smaller Stokes numbers, while  
$H_d/R \sim St^{-1/2} $ and $V_b/R^3 \sim St^{-1}$ for larger Stokes numbers. 
We will now theoretically derive these scaling laws. For small $St$ we follow and extend ref.\ \cite{hic10,man12}: The horizontal length scale 
$L$ of the dimple extension (see figure\ \ref{fig1}a) follows from geometrical arguments as $L \sim \sqrt{H_d R}$, and
$u_r$ from mass conservation as $u_r \sim U L/H_d$. The Stokes equation (\ref{stokes}) suggests
$P_g \sim L \eta_g u_r/H_d^2$ as estimate for the gas pressure below the falling drop at touch-down. The liquid pressure $P_l$
can be estimated from the unsteady Bernoulli equation: dimensional analysis for the deceleration timescale $H_d/U$ 
and for the potential in the liquid $\sim UL$, resulting in $P_l\sim \rho_l U^2 L/H_d$. Since the liquid drop will be deformed when $P_g \sim P_l$, one finally obtains the scaling for the dimple height and the bubble volume:
\be 
H_d \sim R St^{2/3}, \qquad V_b \sim L^2 H_d \sim R^3 St^{4/3}.
\label{inertial-scaling}
\ee
This describes the air bubble in the inertial regime, i.e. large impact velocities, in agreement with our  experimental and numerical findings. 

For large $St$, corresponding to small impact velocity and small droplet radius, capillarity will take over and try to smoothen the dimple
out. Then the right hand side of the Stokes equation (\ref{stokes}) must be balanced with the Laplace pressure 
$\gamma  \kappa$, where $\kappa \sim H_d/L^2$ is the curvature of the dimple. Using once more that the gas pressure $P_g \sim L \eta_g u_r/H_d^2$, one immediately obtains
\be
\begin{array}{l}
\displaystyle {H_d\over R}  \sim \sqrt{Ca} \sim \sqrt{St We} \sim {\eta_g \over \sqrt{\gamma \rho_l R}} St^{-1/2}, \\
\displaystyle {V_b\over R^3}  \sim  {\eta_g^2 \over \gamma \rho_l R} St^{-1}, \\
\end{array} 
\label{cap-scaling}
\ee
as scaling in the capillary regime. Again, this agrees well with the experimental and numerical findings. The crossover between the regimes,
corresponding to the maximal air bubble entrainment, occurs at 
\be St_o \sim Ca_o^{3/4} \quad \hbox{or}\quad U_o \sim {\eta_g^{1/7} \gamma^{3/7} \over \rho_l^{4/7} R^{4/7}} .
\label{opt}
\ee 
Using prefactors obtained from our experimental data in figure \ref{fig2}, for an ethanol droplet of 0.9$\,$mm radius, this translates to 
an impact velocity $U_o $ of 0.25$\,$m/s.
What is the physical reason for the maximum? For higher velocities inertia dominates and flattens the droplet at impact. For lower velocities and/or smaller droplets the capillary forces try to keep the drop spherical. In between these two regimes the maximal air entrainment under the droplet is achieved.

For many applications air entrainment is undesirable and maximal wetting must be achieved. This holds for immersion lithography, wafer drying, glueing, agricultural applications \cite{swi05,Winkels2011}. Intriguingly, for inkjet drops of radius $R \sim 10\mu$m, the optimal velocity according to (\ref{opt}) is approximately 1 m/s. This lies exactly in the range at which inkjet usually operates (typically a few m/s), and relatively large bubbles will thus be entrapped \cite{dam04}. For immersion lithography the entrapment of even micron-sized bubbles can cause practical limitations \cite{swi05,Winkels2011}. This technology is based on optical imaging of nanoscale structures, for which the optics is immersed in water to push the limits of spatial resolution. Clearly, it is crucial to avoid bubbles or to minimize their size, which also has bearing in cleaning and drying of wafers. Ideally, one should stay as far as possible from the optimal air entrainment impact velocity. Our findings will help to achieve this goal and thus optimal wetting.

\noindent 
{\it Acknowledgements:} We gratefully acknowledge helpful discussions with Christophe Clanet.
This study was financially supported by NWO, FOM, an ERC advanced grant, and ASML.


\end{document}